\documentclass[runningheads, a4paper, 10pt]{llncs}
\usepackage{graphicx}
\usepackage{caption}
\usepackage[hyphens]{url}
\usepackage{xspace}
\usepackage{authblk}
\usepackage{multirow}
\usepackage{longtable}
\usepackage[normalem]{ulem}
\usepackage{listings}
\usepackage{xcolor}
\usepackage[utf8]{inputenc}
\definecolor{mygreen}{rgb}{0,0.6,0}
\definecolor{mygray}{rgb}{0.5,0.5,0.5}
\definecolor{mymauve}{rgb}{0.58,0,0.82}
\definecolor{base03}{HTML}{002b36}
\definecolor{base02}{HTML}{073642}
\definecolor{base01}{HTML}{586e75}
\definecolor{base00}{HTML}{657b83}
\definecolor{base0}{HTML}{839496}
\definecolor{base1}{HTML}{93a1a1}
\definecolor{base2}{HTML}{eee8d5}
\definecolor{base3}{HTML}{fdf6e3}
\definecolor{yellow}{HTML}{b58900}
\definecolor{orange}{HTML}{cb4b16}
\definecolor{red}{HTML}{dc322f}
\definecolor{magenta}{HTML}{d33682}
\definecolor{violet}{HTML}{6c71c4}
\definecolor{blue}{HTML}{268bd2}
\definecolor{cyan}{HTML}{2aa198}
\definecolor{green}{HTML}{859900}
\newcommand{\codefont}{\fontfamily{pcr}\selectfont}

\lstdefinelanguage{JavaScript}{
  keywords={function, var, const, class, default, typeof, void, export, extends, import, super},
  keywordstyle=\color{blue}\bfseries,
  ndkeywords={break, case, try, catch, continue, debugger, delete, do, else, export, throw, while,
  with, yield, for, finally, if,  in, instanceof, new, return, switch, null},
  ndkeywordstyle=\color{yellow}\bfseries,
  identifierstyle=\color{base02},
  sensitive=false,
  comment=[l]{//},
  morecomment=[s]{/*}{*/},
  commentstyle=\color{base1},
  stringstyle=\color{cyan}\bfseries,
  morestring=[b]',
  morestring=[b]"
}

\lstdefinelanguage{HTML}{
  keywords={a, abbr, address, area, article, aside, audio, b, base, bdi,
  bdo, blockquote, body, br, button, canvas, caption, cite, code, col, colgroup, datalist, dd, del,
  details, dfn, dialog, div, dl, dt, em, embed, fieldset, figcaption, figure, footer, form, h1, h2, h3, h4, h5, h6, head, header, hr, html, i, iframe, img, input, ins, input, ins, kbd, keygen,
  label, legend, li, link, main, map, mark, menu, menuitem, meta, meter, nav, noscript, object, ol, optgroup, option, output, p, param, picture, pre, progress, q, rp, rt, ruby, s, samp, script, section, select, small, source, span, strong, style, sub, summary, sup, table, tbody, td,
  textarea, tffot, th, thead, time, title, tr, track, u, ul, video, wbr},
  keywordstyle=\color{blue}\bfseries,
  ndkeywords={break, case, catch, class, const, continue, debugger, default, delete, do, else, export, extends, finally, for, function, if, import, in, instanceof, new, return, super, switch, this, throw, try, typeof, var, void, while, with, yield, ||, &&},
  ndkeywordstyle=\color{yellow}\bfseries,
  identifierstyle=\color{base02},
  sensitive=false,
  comment=[l]{//},
  morecomment=[s]{<!--}{-->},
  morecomment=[n]{/*}{*/},
  commentstyle=\color{base1},
  stringstyle=\color{cyan}\bfseries,
  morestring=[b]',
  morestring=[b]"
}

\newcommand{\accesstoken}{\textit{access\_token}\xspace}

\newcommand{\code}{\textit{code}\xspace}

\newcommand{\redirecturi}{\textit{redirect\_uri}\xspace}
\begin{document}
\title{Mitigating CSRF attacks on OAuth 2.0 and OpenID Connect}

\author[1]{Wanpeng Li\inst{1} \and Chris J Mitchell\inst{2} \and Thomas Chen\inst{1}}
\institute{
  Department of Electrical \& Electronic Engineering\\
  City, University of London, UK  \authorcr {\tt \{Wanpeng.Li, Tom.Chen.1\}@city.ac.uk}
  \and
  Information Security Group\\
  Royal Holloway,
  University of London, UK  \authorcr {\tt C.Mitchell@rhul.ac.uk}
}

\maketitle

\begin{abstract}
Many millions of users routinely use their Google, Facebook and Microsoft
accounts to log in to websites supporting OAuth 2.0 and/or OpenID Connect-based
single sign on. The security of OAuth 2.0 and OpenID Connect is therefore of
critical importance, and it has been widely examined both in theory and in
practice.  Unfortunately, as these studies have shown, real-world
implementations of both schemes are often vulnerable to attack, and in
particular to cross-site request forgery (CSRF) attacks. In this paper we
propose a new technique which can be used to mitigate CSRF attacks against both
OAuth 2.0 and OpenID Connect.

\end{abstract}
\section{Introduction} 
\label{sec:introduction}

Since the OAuth 2.0 authorisation framework was published at the end of 2012 \cite{oauth2}, it has been adopted by a large number of websites worldwide as a means of providing single sign-on (SSO) services. By using OAuth 2.0, websites can reduce the burden of password management for their users, as well as saving users the inconvenience of re-entering attributes that are instead stored by identity providers and provided to relying parties as required.

\begin{figure}[htbp]
 \centering
 \includegraphics[scale=0.4]{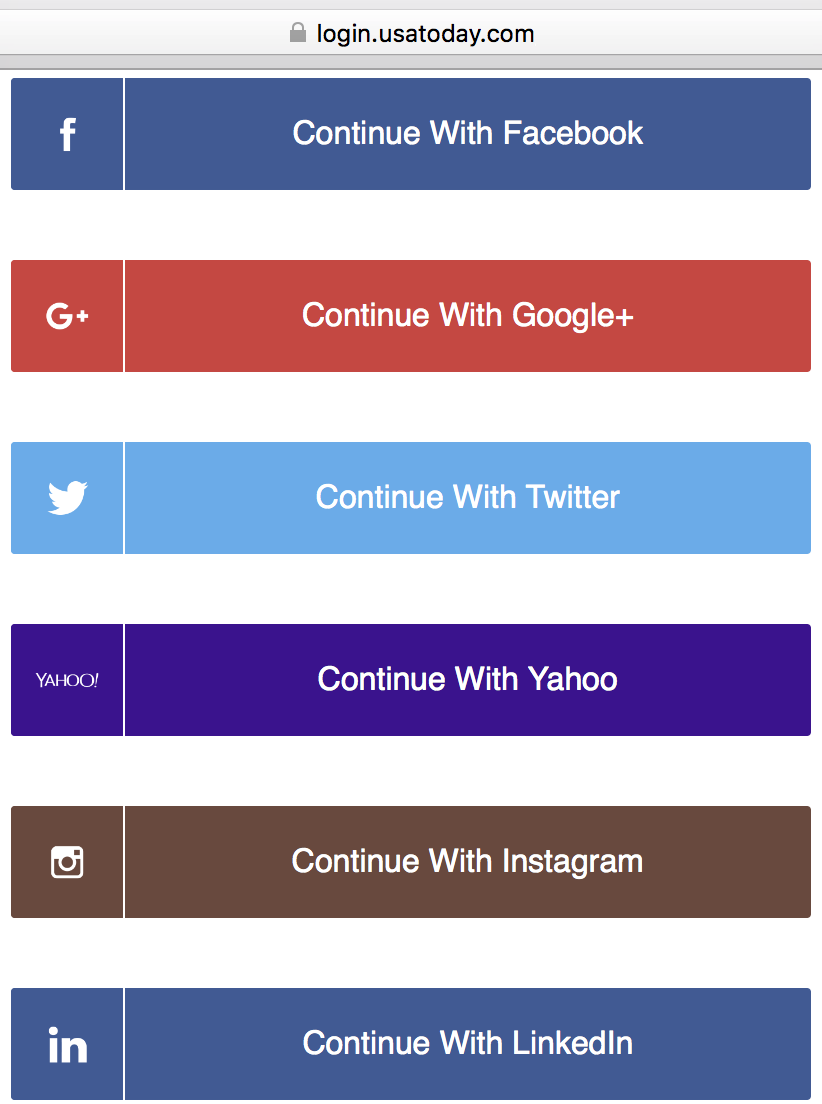}
 \caption{The OAuth 2.0 IdPs supported by USATODAY}
 \label{fig:usatoday}
\end{figure}

There is a correspondingly rich infrastructure of identity
providers (IdPs) providing identity services using OAuth 2.0.
This is demonstrated by the fact that some Relying Parties
(RPs), such as the website
USATODAY\footnote{\url{https://login.usatoday.com/USAT-GUP/authenticate/?}},
support as many as six different IdPs --- see Fig.
\ref{fig:usatoday}.

As discussed in greater detail in section 4, the theoretical security of OAuth 2.0
has been analysed using formal methods
\cite{DBLP:journals/jcs/BansalBDM14,WebSpi,ProVerif,DBLP:journals/iacr/ChariJR11,fett2016comprehensive,pai11,frostig11}. Research focusing on the practical security and privacy properties of
implementations of OAuth 2.0 has also been conducted
\cite{DBLP:conf/ccs/ChenPCTKT14,DBLP:conf/isw/LiM14,wanpeng:dimva2016,DBLP:conf/codaspy/ShehabM14,DBLP:conf/ccs/SunB12,DBLP:conf/sp/WangCW12,DBLP:conf/uss/ZhouE14}.
This latter research has revealed that many real-world implementations of OAuth 2.0 and OpenID Connect have serious vulnerabilities, often because the implementation advice from IdPs is hard to follow.

In this paper we look at a class of cross-site request forgery (CSRF) attacks which have been shown to apply to many real-world implementations of both OAuth 2.0 and OpenID Connect (which is OAuth 2.0-based).  In particular, we examine a new class of mitigations which can be applied to prevent such attacks; such techniques are needed because, although existing mitigations are effective in principle, for a variety of practical reasons these are often not deployed in operational implementations.

The remainder of this paper is structured as follows.
 Section
\ref{sec:background} provides background on OAuth 2.0. In section
\ref{sec:idp_mix_up} we describe implementation strategies that RPs use to
support multiple IdPs. Section \ref{sec:related_work} summarises previous work
analysing the security of real-world OAuth 2.0 implementations. Section
\ref{sec:CSRF} describes the adversary model we use in this paper, and also
gives a detailed description of possible CSRF attacks against OAuth 2.0 and
OpenID Connect. In section \ref{sec:NewApproach} we propose a new approach to
defend against CSRF attacks. In section \ref{sec:using_a_custom_header} we
describe how CSRF attacks can be mitigated for RPs using a specific OAuth 2.0
client library. Section \ref{sec:limitations} describes possible limitations of
our approach and also possible ways of avoiding these limitations. Section
\ref{sec:conclusion} concludes the paper.

\section{Background}
\label{sec:background}

\subsection{OAuth 2.0}
\label{sub:OAuth2}

\begin{figure}[htbp]
 \centering
 \includegraphics[width=0.8\textwidth]{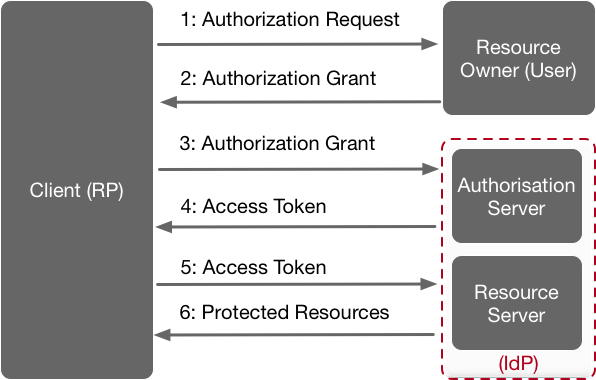}
 \caption{OAuth 2.0 Protocol Flow}
 \label{fig:OAuth2ProcotolFlow}
\end{figure}

The OAuth 2.0 specification \cite{oauth2} describes a system
that allows an application to access resources (typically
personal information) protected by a \emph{resource server} on
behalf of the \emph{resource owner}, through the consumption of
an \emph{access token} issued by an \emph{authorization
server}. In support of this system, the OAuth 2.0 architecture
involves the following four roles (see Fig.
\ref{fig:OAuth2ProcotolFlow}).

\begin{enumerate}

\item The \emph{Resource Owner} is typically an end user.

\item The \emph{Client} is an application running on a
    server, which makes requests on behalf of the resource
    owner (the \emph{Client} is the RP when OAuth 2.0 is
    used for SSO).

\item The \emph{Authorization Server} generates access
    tokens for the client, after authenticating the
    resource owner and obtaining its authorization.

\item The \emph{Resource Server} is a server which stores
    the protected resources and consumes access tokens
    provided by an authorization server (the
    \emph{Resource Server} and \emph{Authorization Server}
    together constitute the IdP when OAuth 2.0 is used for
    SSO).

\end{enumerate}

Fig. \ref{fig:OAuth2ProcotolFlow} provides an overview of the
operation of the OAuth 2.0 protocol. The client initiates the
process by sending (1) an authorization request to the resource
owner. In response, the resource owner generates an
authorization grant (or authorization response) in the form of a \emph{code}, and sends
it (2) to the client. After receiving the authorization grant,
the client initiates an access token request by authenticating
itself to the authorization server and presenting the
authorization grant, i.e.\ the code issued by the resource
owner (3). The authorization server issues (4) an access token
to the client after successfully authenticating the client and
validating the authorization grant. The client makes a
protected source request by presenting the access token to the
resource server (5). Finally, the resource server sends (6) the
protected resources to the client after validating the access
token.

The OAuth 2.0 framework defines four ways for RPs to obtain
access tokens, namely \emph{Authorization Code Grant}, \emph{Implicit Grant},
\emph{Resource Owner Password}, and \emph{Client Credentials Grant}. In this
paper we are only concerned with the \emph{Authorization Code Grant}
and \emph{Implicit Grant} protocol flows. Note that, in the descriptions below, protocol parameters given in bold font are defined as required (i.e.\ mandatory) in the OAuth 2.0 Authorization Framework \cite{oauth2}.

\subsection{OpenID Connect}

OpenID Connect 1.0 \cite{openidconnect} builds an identity
layer on top of the OAuth 2.0 protocol. The added functionality
enables RPs to verify an end user identity by relying on an
authentication process performed by an \emph{OpenID Provider
(OP)}\@.  In order to enable an RP to verify the identity of an
end user, OpenID Connect adds a new type of token to OAuth 2.0,
namely the \emph{id\_token}. This complements the access token
and code, which are already part of OAuth 2.0.  An
\emph{id\_token} contains claims about the authentication of an
end user by an OP together with any other claims requested by
the RP.

OpenID Connect supports three authentication flows
\cite{openidconnect}, i.e.\ ways in which the system
can operate, namely  \emph{Hybrid Flow},
\emph{Authorization Code Flow} and \emph{Implicit Flow}.

\subsection{OAuth 2.0 used for SSO}
\label{sub:OAuth2SSO}

In order to use OAuth 2.0 as the basis of an SSO system, the following role mapping is used:
\begin{itemize}
\item the resource server and authorization server together
    play the IdP role;
\item
the client plays the role of the RP;
\item
the resource owner corresponds to the user.
\end{itemize}

OAuth 2.0 and OpenID Connect SSO systems build on user agent
(UA) redirections, where a user (U) wishes to access services
protected by the RP which consumes the access token generated
by the IdP\@. The UA is typically a web browser. The IdP
provides ways to authenticate the user, asks the user to grant
permission for the RP to access the user's attributes, and
generates an access token on behalf of the user. After
receiving the access token, the RP can access the user's
attributes using the API provided by the IdP\@.

\subsubsection{RP Registration} 
\label{ssub:oauth_resistration}

The RP must register with the IdP before it can use OAuth 2.0. During registration, the IdP gathers security-critical information about the RP, including the RP's redirect URI, i.e.\ \textbf{\textit{redirect\_uri}},
the URI to which the user agent is redirected after the IdP has generated the authorization response and sent it to the RP via the UA\@. As part of registration, the IdP issues the RP with a unique identifier (\textbf{\textit{client\_id}}) and, optionally, a secret (\textit{client\_secret}). If defined, \textit{client\_secret} is used by the IdP to authenticate the RP when using the Authorization Code Grant flow.

\subsubsection{Authorization Code Grant} 
\label{ssub:oauthorization_code_flow}


The OAuth 2.0 Authorization
Code Grant is very similar to the OpenID Connect Authorization Code Flow; for simplicity, we only give the description of the OAuth 2.0 Authorization
Code Grant. This flow relies on certain information having
been established during the registration process, as described in section \ref{ssub:oauth_resistration}. An instance of use of the protocol proceeds as follows.

\begin{enumerate}
\item U $\rightarrow $ RP: The user clicks a login button on the
RP website, as displayed by the UA, which causes the UA to
send an HTTP request to the RP.

\item RP $\rightarrow$ UA: The RP produces an OAuth 2.0
    authorization request and sends it back to the UA. The
    authorization request includes
    \textbf{\textit{client\_id}}, the identifier for the
    client which the RP registered with the IdP
    previously; \textbf{\textit{response\_type=code}},
    indicating that the Authorization Code Grant method is
    requested; \textit{redirect\_uri}, the URI to which the
    IdP will redirect the UA after access has been granted;
    \textit{state}, an opaque value used by the RP to
    maintain state between the request and the callback
    (step 6 below); and \textit{scope}, the scope of the
    requested permission.

\item UA $\rightarrow$ IdP: The UA redirects the request
    which it received in step 2 to the IdP.

\item IdP $\rightarrow$ UA: The IdP first compares the value of
    \textit{redirect\_uri} it received in step 3 (embedded
    in the authorization request) with the registered value;
    if the comparison fails, the process terminates. If the user has already been
    authenticated by the IdP, then the next step is
    skipped. If not, the IdP returns a login form which is
    used to collect the user authentication information.

\item U $\rightarrow$ UA $\rightarrow$ IdP: The user
    completes the login form and grants permission for the
    RP to access the attributes stored by the IdP.

\item IdP $\rightarrow$ UA $\rightarrow$ RP: After (if
    necessary) using the information provided in the login
    form to authenticate the user, the IdP generates an
    authorization response and redirects the UA back to the
    RP\@. The authorization response contains
    \textbf{\textit{code}}, the authorization code
    (representing the authorization grant) generated by the
    IdP; and \textit{state}, the value sent in step 2.

\item RP $\rightarrow$ IdP: The RP produces an access token
    request and sends it to the IdP token endpoint directly
    (i.e.\ not via the UA)\@. The request includes
    \textbf{\textit{grant\_type=authorization\_code}},
    \textbf{\textit{client\_id}}, \textit{client\_secret}
    (if the RP has been issued one),
    \textbf{\textit{code}} (generated in step 6), and the
    \textbf{\textit{redirect\_uri}}.

\item IdP $\rightarrow$ RP: The IdP checks
    \textbf{\textit{client\_id}}, \textit{client\_secret} (if present),  \textbf{\textit{code}} and
    \textbf{\textit{redirect\_uri}} and, if the checks succeed, responds to the RP
    with \textit{access\_token}.

\item RP $\rightarrow$ IdP: The RP passes
    \textit{access\_token} to the IdP via a defined API to
    request the user attributes.

\item IdP $\rightarrow$ RP: The IdP checks
    \emph{access\_token} (how this works is not specified in the OAuth 2.0 specification) and, if
    satisfied, sends the requested user attributes to the
    RP\@.
\end{enumerate}

\subsubsection{Implicit Grant}
\label{ssub:implicit_grant}
The OAuth 2.0 Implicit Grant is very similar to the OpenID Connect Implicit Flow and Hybrid Flow; for simplicity, we only give the description of the OAuth 2.0 Implicit Grant. This flow has a similar sequence of
steps to Authorization Code Grant. We specify below only those
steps where the Implicit Grant flow differs from the
Authorization Code Grant flow.

\begin{enumerate}

\item[2.] RP $\rightarrow$ UA: The RP produces an OAuth 2.0
    authorization request and sends it back to the UA\@.
    The authorization request includes
    \textbf{\textit{client\_id}}, the identifier for the
    client which the RP registered with the IdP
    previously;  \textbf{\textit{response\_type=token}},
    indicating that the Implicit Grant  is requested;
    \textit{redirect\_uri}, the URI to which the IdP will
    redirect the UA after access has been granted;
    \textit{state}, an opaque value used by the RP to
    maintain state between the request and the callback
    (step 6 below); and \textit{scope}, the scope of the
    requested permission.

\item[6.] IdP $\rightarrow$ UA $\rightarrow$ RP: After (if
    necessary) using the information provided in the login
    form to authenticate the user, the IdP generates an
    access token and redirects the UA back to the RP using the value of
   \textit{redirect\_uri} provided in step 2. The access token
    is appended to \textit{redirect\_uri} as a URI fragment
    (i.e.\ as a suffix to the URI following a \# symbol).

\end{enumerate}

As URI fragments are not sent in HTTP requests, the access token is
not immediately transferred when the UA is redirected to the
RP\@. Instead, the RP returns a web page (typically an HTML
document with an embedded script) capable of accessing the full
redirection URI, including the fragment retained by the UA, and
extracting the access token (and other parameters) contained in
the fragment; the retrieved access token is returned to the
RP\@. The RP can now use this access token to retrieve data
stored at the IdP\@.

\section{Supporting multiple IdPs} 
\label{sec:idp_mix_up}

As described in section \ref{sec:introduction}, many RPs
support more than one IdP\@.  This recognises the fact that
users will have trust relationships with varying sets of IdPs
--- for example, one user may prefer to trust Facebook, whereas another may
prefer Google.

In this section we describe two ways in which this is achieved in practice.

\subsection{Using redirect URIs}
\label{sub:using_redirect_URIs}

One way in which an RP can support multiple IdPs is to
register a different \textit{redirect\_uri} with each IdP, and
to set up a sign-in endpoint for each. The can then use the
endpoint on which it receives an authorization response to
recognise which IdP sent it. For example,
AddThis\footnote{\url{http://www.addthis.com/}} has registered the URIs:
\begin{itemize}
\item
    \url{https://www.addthis.com/darkseid/account/register-facebook-return}
    as its \redirecturi for Facebook, and
\item
    \url{https://www.addthis.com/darkseid/account/register-google-return}
    as its \redirecturi for Google.
\end{itemize}
If AddThis receives an authorization response at the endpoint
    \url{https://www.addthis.com/darkseid/account/register-facebook-return?code=[code_generated_by_Facebook]},
(in step 7 of section~\ref{ssub:oauthorization_code_flow}), it
assumes that this response was generated by Facebook, and thus
sends the authorization \textit{code} to the Facebook server
(step 8 of section~\ref{ssub:oauthorization_code_flow}) to
request an \emph{access\_token}.

\subsection{Explicit User Intention Tracking}
\label{sub:explicit_user_itention_tracking}

Registering a different redirection URI for each IdP is not the
only approach that could be used by an RP to support multiple
IdPs.  If an RP does not register a different redirection URI
for each IdP, it can instead keep a record of the IdP each user wishes to use to authenticate (e.g.\ it could
save the identity of the user's selected IdP to a cookie).

In this case, when a authorization response is received by the
RP, the RP can retrieve the identity of the IdP from the cookie
and then send the \emph{code} to this IdP\@. This method is
typically used by RPs that allow for dynamic registration,
where using the same URI is an obvious implementation choice
\cite{fett2016comprehensive}.


\section{Analysing the security of OAuth 2.0 and OpenID Connect}
\label{sec:related_work}

The security properties of OAuth 2.0 have been analysed using formal methods. Pai et al.\
\cite{pai11} confirmed a security issue described in the OAuth
2.0 Thread Model \cite{oauth2threat} using the Alloy Framework
\cite{alloy}. Chari et al.\ analysed OAuth 2.0 in the Universal
Composability Security Framework
\cite{DBLP:journals/iacr/ChariJR11} and showed that OAuth 2.0
is secure if all the communications links are SSL-protected.
Frostig and Slack \cite{frostig11} discovered a cross site
request forgery attack in the Implicit Grant flow of OAuth 2.0,
using the Murphi framework \cite{DBLP:conf/cav/Dill96}. Bansal
et al.\ \cite{DBLP:journals/jcs/BansalBDM14} analysed the
security of OAuth 2.0 using the WebSpi \cite{WebSpi} and
ProVerif models \cite{ProVerif}. However, all this work is
based on abstract models, and so delicate implementation
details are ignored.

The security properties of real-world OAuth 2.0 implementations
have also been examined by a number of authors. Wang et al.\
\cite{DBLP:conf/sp/WangCW12} examined deployed SSO systems,
focussing on a logic flaw present in many such systems,
including OpenID\@. In parallel, Sun and Beznosov
\cite{DBLP:conf/ccs/SunB12} also studied deployed OAuth 2.0
systems. Later, Li and Mitchell \cite{DBLP:conf/isw/LiM14}
examined the security of deployed OAuth 2.0 systems providing
services in Chinese. In parallel, Zhou and Evans
\cite{DBLP:conf/uss/ZhouE14} conducted a large scale study of
the security of Facebook's OAuth 2.0 implementation. Chen et
al.\ \cite{DBLP:conf/ccs/ChenPCTKT14}, and Shehab and Mohsen
\cite{DBLP:conf/codaspy/ShehabM14} have looked at the security
of OAuth 2.0 implementations on mobile platforms. Li and
Mitchell \cite{wanpeng:dimva2016} conducted an empirical study
of the security of the OpenID Connect-based SSO service
provided by Google.

These latter studies have revealed a number of vulnerabilities arising in implementations of these systems.  Of particular interest here are vulnerabilities which can lead to CSRF attacks, and we next describe how these attacks can arise.


\section{CSRF Attacks against OAuth 2.0 and OpenID Connect} 
\label{sec:CSRF}

\subsection{Adversary Model}
\label{sub:AttackModel}

We suppose that the adversary has the capabilities of a {\bf
web attacker}, i.e.\ it can share malicious links or post
comments which contain malicious content (e.g.\ stylesheets or
images) on a benign website, and/or can exploit vulnerabilities
in an RP website. The malicious content might trigger the web
browser to send an HTTP/HTTPS request to an RP and IdP using
either the GET or POST methods, or execute JavaScript scripts
crafted by the attacker.

\subsection{CSRF attacks} 
\label{ssub:csrf_attack}

A cross site request forgery (CSRF) attack
\cite{DBLP:conf/ccs/BarthJM08,
burns2005cross,DBLP:conf/esorics/RyckDJP11,
jovanovic2006preventing, DBLP:conf/fc/MaoLM09,
DBLP:conf/issre/ShahriarZ10,zeller2008cross} operates in the
context of an ongoing interaction between a target web browser
(running on behalf of a target user) and a target website. The
attack involves a malicious website causing the target web
browser to initiate a request of the attacker's choice to the
target website. This can cause the target website to execute
actions without the involvement of the user. In particular, if
the target user is currently logged into the target website,
the target web browser will send cookies containing an
authentication token generated by the target website for the
target user, along with the attacker-supplied request, to the
target website. The target website will then process the
malicious request as through it was initiated by the target
user.

There are various ways in which the target browser could be made to send the spurious request. For example, a malicious website visited by the browser could use the HTML $<$img$>$ tag's src attribute to specify the malicious request URL, which will cause the browser to silently use a GET method to send the request to the target website.

According to the OWASP Top 10 -- 2013 report \cite{owasp_2013_top10} released by the Open Web Application Security Project (OWASP) in 2013, the CSRF attack is ranked as No.\ 8  in the 10 most critical web application security risks.  This means that such attacks represent a real danger in practice.


\subsection{CSRF Attacks Against the Redirect URI}
\label{CH:Issues:SecurityandPrivacyIssues:Abstract:CSRF}

CSRF attacks
against the OAuth 2.0 \emph{redirect\_uri} \cite{oauth2threat} can allow an attacker to obtain authorization to access OAuth-protected resources without the consent of the user. Such an attack is possible for both the Authorization Code Grant Flow and the Implicit Grant Flow.

An attacker first acquires a \textit{code} or \accesstoken relating to its own protected resources. The attacker then aborts the redirect flow back to the RP on the attacker's own device, and then, by some means, tricks the victim into executing the redirect back to the RP\@. The RP receives the redirect, fetches the attributes from the IdP, and associates the victim's RP session with the attacker's resources that are accessible using the tokens. The victim user then accesses resources on behalf of the attacker.

The impact of such an attack depends on the type of resource accessed.  For example, the user might upload private data to the RP, thinking it is uploading information to its own profile at this RP, and this data will subsequently be available to the attacker; as described by Li and Mitchell \cite{DBLP:conf/isw/LiM14}, an attacker can use a CSRF attack to control a victim user's RP account without knowing the user's username and password.

\subsection{Existing CSRF Defences}
\label{sub:ExistingCSRF}
Barth et al.\ \cite{DBLP:conf/ccs/BarthJM08} describe four mechanisms that a website can use to defend itself against CSRF attacks: validating a secret token, validating the HTTP Referer header, including additional headers with an XMLHttpRequest, and validating the HTTP Origin header. All of these mechanisms are in use on the web today, but none of them are entirely satisfactory.
\begin{itemize}
  \item \textbf{Secret Validation Token.} One approach to defending against CSRF attacks is to send additional information in the form of a secret validation token in each HTTP request; this token can be used to determine whether the request came from an authorized source. The ``validation token'' should be hard to guess for attacker who does not already have access to the user's account. If a request does not contain a validation token, or the token does not match the expected value, the server should reject the request.
  \item \textbf{The Referer Header.} In many cases, when the browser issues an HTTP request, it includes a Referer header that indicates which URL initiated the request. The Referer header, if present, distinguishes a same-site request from a cross-site request because the header contains the URL of the site making the request. A site can defend itself against cross-site request forgery attacks by checking whether the request in question was issued by the site itself. However, the Referer might contain sensitive information that impinges on the privacy of web users.
  \item \textbf{Custom HTTP Headers.} Custom HTTP headers can be used to
      prevent CSRF attacks, because browsers prevent sites from sending
      custom HTTP headers to another site but allow sites to send custom
      HTTP headers to themselves using XMLHttpRequest. For example, the
      prototype.js JavaScript
      library\footnote{\url{http://prototypejs.org}} uses this approach and
      attaches the X-Requested-With header with the value XMLHttpRequest.
      To use custom headers as a CSRF defence, a site must issue all
      state-modifying requests using XMLHttpRequest, attach the custom
      header (e.g.\ X-Requested-with), and reject all state-modifying
      requests that are not accompanied by the header. This works because
      XMLHttpRequest does not allow an attacker to make a request to a
      third party domain by default. Thus, it is not possible for an
      attacker to forge a request with a spoofed X-Requested-With header.
  \item \textbf{The Origin Header}. In many cases, the browser includes an Origin header, which indicates the request originates from an HTTP POST request as well as a Cross-Origin Resource Sharing (CORS)\footnote{\url{https://developer.mozilla.org/en-US/docs/Glossary/CORS}} request. Its use is similar to the Referer header, but it does not include any path information, only the server name.

\end{itemize}

\subsection{Existing CSRF Defences for OAuth 2.0}

As described in the OAuth Threat model \cite{oauth2threat}, two possible
mitigations \cite{oauth2threat} for a CSRF attack are as follows.
\begin{itemize}
    \item A \textit{state} parameter should be used to link the authorization request to the redirect URI used to deliver the \code or \accesstoken (see Secret Validation Token in section \ref{sub:ExistingCSRF}).
    \item RP developers and end users should be educated not to follow untrusted URLs.
\end{itemize}

\label{sub:CSRF_real}

The OAuth Threat model does not recommend use of the three other CSRF defences
described in section \ref{sub:ExistingCSRF}. This is because the OAuth 2.0
response to the redirect URI is a cross-site request, since the request is
generated by the IdP and is redirected to the RP by the browser; all the
existing CSRF defences (except for use of the secret token) will therefore not
work in this case.

It is important to observe that both the recommended mitigations delegate the responsibility for correctly implementing CSRF countermeasures to the RP developers. However, in practice, RPs do not always implement CSRF
countermeasures in the recommended way. A study conducted by
Shernan et al.\ \cite{DBLP:conf/dimva/ShernanCTTB15} in 2015
found that 25\% of websites in the Alexa Top 10,000
domains using Facebook's OAuth 2.0 service appear vulnerable
to CSRF attacks. Further, a 2016 study conducted by Yang et
al.\ \cite{DBLP:conf/ccs/YangLLZH16} revealed that 61\% of
405 websites using OAuth 2.0 (chosen from the 500 top-ranked US and Chinese
websites) did not implement CSRF
countermeasures; even worse, for those RPs which support the \textit{state} parameter, 55\% of them are still vulnerable to CSRF attacks due to  misuse/mishandling of the \textit{state} parameter. They also disclosed four scenarios where the \textit{state} parameter can be misused by the RP developers.

This means that, if CSRF attacks are to be prevented in practice, new and simple-to-implement CSRF countermeasures would be extremely valuable.  This motivates the work described in the remainder of this paper.

\section{A new approach}


\label{sec:NewApproach}

Since the requirement to add a state parameter to an authorization request is
often ignored by RP developers, large numbers of real-world OAuth 2.0
implementations are vulnerable to CSRF attacks; moreover traditional Referer
header, Origin header and Custom header countermeasures are infeasible in the
OAuth 2.0 framework (see section \ref{sub:CSRF_real}). We propose to combine
the Referer header and the fact that RPs register different URIs for different
IdPs (see section \ref{sub:using_redirect_URIs}) to provide a novel means of
mitigating CSRF attacks. We first describe how a Referer header can be used to
mitigate CSRF attacks against the \redirecturi in both the Authorization Code
Grant Flow of OAuth 2.0 and the (very similar) Authorization Code Flow of
OpenID Connect.

\subsection{Protecting the Authorization Code (Grant) Flow}  \label{code_protection}

\begin{figure}[htbp]
 \centering
 \includegraphics[scale=0.5]{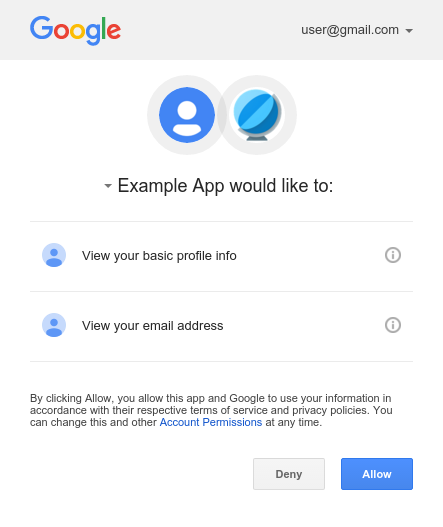}
 \caption{An Google Authorization Grant Example}
 \label{fig:googlegrant}
\end{figure}

Normally an authorization response is only generated after a user clicks on a
grant button rendered by the IdP (see fig. \ref{fig:googlegrant}). The HTTP
message (see, for example, listing \ref{listing:HTTPMessage}) of such an
authorization response contains a Referer header which points to the IdP
domain.

\begin{lstlisting}[caption={HTTP message of a normal OAuth 2.0 Authorization Response}, label={listing:HTTPMessage}]
// HTTP message of a normal OAuth 2.0 Authorization response generated by AIdP for RP.com
// privacy relerated values are trimmed using ***

GET /AIdP-callback?code=[code_generated_by_AIdP]
Host: RP.com
User-Agent: ***
Accept: ***
Accept-Language: en-US,en;q=0.5
Referer: https://AIdP.com/
Cookie: ***
Connection: close
\end{lstlisting}

In practice, major IdPs, such as Google, Facebook and Microsoft, implement an
‘automatic authorization granting’ feature \cite{wanpeng:dimva2016}. That is,
when the user has logged in to his/her OAuth 2.0 IdP account, the IdP generates
an authorization response without explicit user consent. The HTTP message (see,
for example, listing \ref{listing:autogrant}) of such an authorization response
contains a Referer header which points to the RP domain.

The proposed mitigation operates as follows.  When the RP receives an
authorization response, it first retrieve the identity of the IdP from
\redirecturi and then checks that the domain in the Referer header is either
the RP Domain or the IdP domain. If the domain of the Referer header is one of
these two values, then the RP knows this is a genuine authorization response
coming from the IdP; otherwise, the RP should discard this HTTP message and
send an error page to the user.

\begin{lstlisting}[caption={HTTP message of an automatic authorization granting OAuth 2.0 Authorization Response}, label={listing:autogrant}]
// HTTP message of an automatic authorization granting OAuth 2.0 Authorization response generated by AIdP for RP.com
// privacy relerated values are trimmed using ***

GET /AIdP-callback?code=[code_generated_by_AIdP]
Host: RP.com
User-Agent: ***
Accept: ***
Accept-Language: en-US,en;q=0.5
Referer: https://RP.com/
Cookie: ***
Connection: close
\end{lstlisting}

As an example of how this might work in practice, suppose a web attacker puts a link \url{https://rp.com/AIdP-callback?code=[code_belongs_to_attacker_generated_by_AIdP]} on \url{attacker.com} to try to launch a CSRF attack against the
\redirecturi RP registered with AIdP\@. The HTTP message of the attack request will contain a Referer header which points to \url{attacker.com} (see listing \ref{listing:CSRFAttack}). The RP is able to detect this is an attack message by comparing the identity of the IdP or its own domain with the domain in the Referer header.

\begin{lstlisting}[caption={HTTP message of a CSRF attack against \textit{redirect\_uri}}, label={listing:CSRFAttack}]
// HTTP message of a normal OAuth 2.0 Authorization response generated by AIdP for RP.com
// privacy relerated values are trimmed using ***

GET /AIdP-callback?code=[code_belongs_to_attacker_generated_by_AIdP]
Host: RP.com
User-Agent: ***
Accept: ***
Accept-Language: en-US,en;q=0.5
Referer: https://attacker.com/
Cookie: ***
Connection: close

\end{lstlisting}

\subsection{Protecting the Implicit Grant Flow}

We next describe how a Referer header can be used to mitigate CSRF attacks against the \redirecturi in the Implicit Grant Flow of OAuth 2.0 or the Implicit Flow and Hybrid Flow of OpenID Connect.


As an example of how this might work in practice, suppose a web attacker
creates the link
\url{https://rp.com/AIdP-callback#access_token=[accsstoken_belongs_to_attacker_generated_by_AIdP]}
on \url{attacker.com} to try to launch a CSRF attack against the \redirecturi
RP registered with AIdP\@. As discussed in section \ref{ssub:implicit_grant},
the \accesstoken in the Implicit Grant Flow is not immediately transferred when
the UA is redirected to the RP\@. Thus the HTTP request message looks similar
to the CSRF HTTP request described in listing \ref{listing:csrfImplicit}. Note
that the only difference between a normal HTTP message and a CSRF HTTP message
in the OAuth 2.0 Implicit Grant is the Referer header. The RP can detect a CSRF
attack by checking the domain of the Referer header is either the IdP identity
it retrieves from the \redirecturi or its own domain; it can then respond with
different HTML depending on the HTTP messages it has received (see lines 22 and
47 in listing \ref{listing:csrfImplicit}).

\begin{lstlisting}[caption={Preventing CSRF attacks on OAuth 2.0 Implicit Grant}, label={listing:csrfImplicit}, language=JavaScript]

// A normal HTTP request message to the RP redirect_uri in the OAuth 2.0 Implicit Grant

GET /AIdP-callback
Host: RP.com
User-Agent: ***
Accept: ***
Accept-Language: en-US,en;q=0.5
Referer: https://AIdP.com
Cookie: ***

// The HTTP response message
HTTP/1.1 200 OK
Date: ***
Server: ***
Last-Modified: ***
Content-Length: ***
Content-Type: text/html
Connection: Closed

<html>
<body>
<h1>This HTML can be used to extract the access_token!</h1>
......
</body>
</html>

// The HTTP request message of a CSRF attack on the RP redirect_uri in OAuth 2.0 Implicit Grant

GET /AIdP-callback
Host: RP.com
User-Agent: ***
Accept: ***
Accept-Language: en-US,en;q=0.5
Referer: https://attacker.com
Cookie: ***

// The HTTP response message
HTTP/1.1 200 OK
Date: ***
Server: ***
Last-Modified: ***
Content-Length: ***
Content-Type: text/html
Connection: Closed

<html> <body> <h1>A CSRF attack is detected on the AIdP signin
endpoint!</h1> </body> </html>
\end{lstlisting}


\subsection{Supporting multiple IdPs}

In the previous sections we described how the Referer header can be used to
mitigate CSRF attacks against RPs which use \redirecturi to track the user
login intention. We now describe how the Referer header can be used to protect
RPs using explicit user intention tracking.

The user log-in intention is stored in the session (Jsession=12345 in the
example below).  Thus, when the RP receives an authorization response (such as
that in listing \ref{listing:Usertracking}), it retrieves the IdP's identity
from the session and checks whether the domain of the Referer header is either
the IdP identity or its own domain. If the two values are the same then the RP
knows that this is a genuine authorization response; otherwise, it should
respond to the user with an error page.

\begin{lstlisting}[caption={HTTP message of a CSRF attack against \textit{redirect\_uri}}, label={listing:Usertracking}]
// HTTP message of a normal OAuth 2.0 Authorization response generated by AIdP for RP.com using explicit user intention tracking
// privacy relerated values are trimmed using ***

GET /oauth2-callback?code=[code_belongs_to_attacker_generated_by_AIdP]
Host: RP.com
User-Agent: ***
Accept: ***
Accept-Language: en-US,en;q=0.5
Referer: https://AIdP.com/
Cookie: Jsession=12345
Connection: close
\end{lstlisting}

\section{Defending RPs using specific IdP client libraries} 
\label{sec:using_a_custom_header}

Many IdPs, such as
Facebook\footnote{\url{https://developers.facebook.com/docs/facebook-login/web}}
and
Google\footnote{\url{https://developers.google.com/identity/sign-in/web/devconsole-project}},
implement their own OAuth 2.0 client libraries. RPs can use these libraries to
simplify integration of the Facebook and Google OAuth 2.0 services with their
websites. These libraries use postMessage
\cite{OpenIDConnectSessionManagement2014} to deliver OAuth 2.0 responses to the
RP client. The RP client must then use XMLHttpRequest to send the OAuth 2.0
response back to the RP OAuth 2.0 callback endpoint, e.g.\
\url{https://www.rp.com/AIdP-callback}.

The RP OAuth 2.0 callback endpoint might be different from the \redirecturi the RP registered with the IdP\@, e.g.\ \url{https://www.rp.com}; for example, Google requires RPs to register an \textit{origin} value if they want to use the Google OAuth 2.0 client libraries. Because the request to the RP's OAuth 2.0 callback endpoint is initiated from the RP client using XMLHttpRequest, the Referer header in the HTTP message of the request always points to the RP domain.

An RP using these client libraries can detect a CSRF attack by checking that
the domain in the Referer header of the HTTP message (i.e.\ \url{RP.com} in the
example given in listing \ref{listing:Client} ).

\begin{lstlisting}[caption={Defending RPs using specific IdP client libraries}, label={listing:Client}, language=JavaScript]
// HTTP message of the request to the RP's OAuth 2.0 callback endpoint

GET /AIdP-callback?code=[code_generated_by_AIdP]
Host: RP.com
User-Agent: ***
Accept: ***
Accept-Language: en-US,en;q=0.5
Referer: https://RP.com
Cookie: ***
Connection: close
\end{lstlisting}


\section{Limitations of our approach}
\label{sec:limitations}

One possible limitation of the Referer header approach described above is that
a UA might compress the Referer header in a (non-secure) HTTP request if the
referring page is transferred via a secure protocol (e.g.\ HTTPS) \cite{http}.
This means that an RP which uses HTTP to register its \redirecturi with an IdP
cannot use the approach described in section \ref{sec:NewApproach} to defend
against CSRF attacks against its \redirecturi, since as part of compression the
Referer header will be removed by the web browser when it redirects the
authorization response to the RP (note that we assume here IdPs use HTTPS at
their OAuth 2.0 authorization endpoint).

This limitation is negated by the fact that that some IdPs,
such as Amazon and Microsoft, require the RP to register its
\redirecturi using the HTTPS protocol. This means that the
attack mitigation described above will work successfully for
RPs supporting Amazon and Microsoft login. It would clearly be
beneficial if other IdPs could change their registration
process to require RPs to register their \redirecturi using
HTTPS, enabling all RPs to use our approach to mitigate CSRF
attacks.

\section{Conclusions} 
\label{sec:conclusion}

In this paper, we have proposed a new class of mitigations
which can be applied to prevent CSRF attacks against
\redirecturi in OAuth 2.0 and OpenID Connect. Our approach is
practical and simple to implement, and requires no changes to
the IdP service; i.e.\ it can be adopted by an RP independently
of what any other party does. We hope that RPs can adopt this
approach to provide an additional layer of protection against
CSRF attacks for their OAuth 2.0 and/or OpenID Connect
services.  Of course, adoption would likely be increased if
this measure was recommended by major IdPs and/or included in
the relevant specifications.

\bibliographystyle{plain}
\bibliography{bibliography}

\end{document}